%% file: template.tex
\documentclass{Interspeech}

\interspeechcameraready

\title{Multivariate Probabilistic Assessment of Speech Quality}

\author[affiliation={1}]{Fredrik}{Cumlin}
\author[affiliation={2}]{Xinyu}{Liang}
\author[affiliation={3}]{Victor}{Ungureanu}
\author[affiliation={3}]{Chandan K.A.}{Reddy}
\author[affiliation={3}]{Christian}{Schüldt}
\author[affiliation={1}]{Saikat}{Chatterjee}

\affiliation{Department of Information Science and Engineering}{KTH Royal Institute of Technology}{Sweden}
\affiliation{}{HCLTech}{Sweden}
\affiliation{}{Google LLC}{USA}
\email{\{fcumlin, sach\}@kth.se, hopeliang@icloud.com, \{ungureanu, chandanka, cschuldt\}@google.com}

\usepackage{comment}
\usepackage{cite}
\usepackage{amsmath,amssymb,amsfonts}
\usepackage{algorithmic}
\usepackage{graphicx}
\usepackage{textcomp}
\usepackage{xcolor}
\usepackage{subcaption}
\usepackage{adjustbox}

\begin{document}
\ninept

\maketitle

\begin{abstract}
    

    The mean opinion score (MOS) is a standard metric for assessing speech quality, but its singular focus fails to identify specific distortions when low scores are observed. The NISQA dataset addresses this limitation by providing ratings across four additional dimensions: noisiness, coloration, discontinuity, and loudness, alongside MOS. In this paper, we extend the explored univariate MOS estimation to a multivariate framework by modeling these dimensions jointly using a multivariate Gaussian distribution. Our approach utilizes Cholesky decomposition to predict covariances without imposing restrictive assumptions and extends probabilistic affine transformations to a multivariate context. Experimental results show that our model performs on par with state-of-the-art methods in point estimation, while uniquely providing uncertainty and correlation estimates across speech quality dimensions. This enables better diagnosis of poor speech quality and informs targeted improvements.

\end{abstract}

\section{Introduction}

Speech quality assessment (SQA) is the task of assessing the speech quality of speech clips. Subjective SQA is done by letting several raters rate the quality of the speech clip based on an ordinal scale such as the mean-opinion-score (MOS) scale \cite{MOSbook}. Subjective SQA is considered the paramount method for evaluating speech quality, but is hindered by cost and time requirements. Automating the assessment framework with objective methods is thus desirable. 

The objective non-intrusive SQA task is to estimate the quality of a distorted speech clip without a reference signal. Several deep neural network (DNN)-based methods have been developed recently for this task. Most provide a point estimate for the speech quality, and three recent works, DeePMOS-$\mathcal{N}$, DeePMOS-$\mathcal{B}$ and DNSMOS Pro \cite{DeePMOS, DeePMOS-B, DNSMOSp}, predict a one-dimensional posterior distribution for deeper insights.

Following the probabilistic works, our objective is to predict a multivariate posterior distribution in the multivariate speech quality assessment task, where several speech quality labels are given to the speech clips. An example includes the ITU-T P.835 subjective assessment framework \cite{p835}, where three quality values are given measuring the overall quality, the background noise quality, and the speech signal quality. A multivariate distribution can help assess causes for degradation with quantified correlations, which can help find preventative measures to improve the overall speech quality.

Based on the MOSNet architecture \cite{MOSNet}, several objective non-intrusive DNN-based end-to-end methods have been developed. The architectural design can be described by a convolutional neural network (CNN) followed by a bi-directional long short-term memory (BLSTM). Examples include LDNet and LaMOSNet \cite{LDNet, LaMOSNet}. DNSMOS and DNSMOS Pro use another design where a global maximum pooling layer was used for temporal considerations \cite{DNSMOS, DNSMOSp}.

Recently, more interest has been given to self-supervised learning (SSL) techniques due to improved performance and robustness over end-to-end methods. Examples include UTMOS, ZevoMOS, and LE-SSL-MOS \cite{UTMOS, ZevoMOS, LE-SSL-MOS}. Herein, a pre-trained feature extractor, such as Hubert and Wav2vec2.0 \cite{Hubert, w2v2}, is used to extract relevant features. A head is then trained using the features as input to target MOS. Experiments provide evidence of improved generalization ability \cite{SSL-MOS}, but at the expense of much increased computational cost.

The abovementioned methods are only trained to predict a distribution or point estimate of a scalar value of MOS. DNSMOS (p835 variant) \cite{dnsmos_p835}, NISQA \cite{NISQA}, and Conformer \cite{probing} have extended the 1-dimensional case to multidimensional quality prediction. All methods predict point estimates.

\textbf{Our contributions:} In this work we have designed a learning algorithm that predicts posterior multivariate Gaussian distribution of quality given a speech clip when several speech quality aspects are considered. The contributions are as follows:
\begin{itemize}
    \item We design an architecture that models a \textit{proper} multivariate Gaussian posterior, using the Cholesky decomposition.
    \item We generalize the maximum likelihood training framework of \cite{DNSMOSp, DeePMOS}, from 1-dimensional Gaussian distributions to multivariate thereof.
    \item We generalize the affine transformation for training unbiased probabilistic estimators in \cite{DNSMOSp} to vector-valued probabilistic predictions.
\end{itemize}

In this study we are not restricted by computational cost hence we can leverage the work of SSL-based approaches. In particular, we use a pre-trained feature extractor and train a prediction head that predicts a multivariate Gaussian distribution.

\section{Method}
\label{sec:method}

\subsection{Problem formulation}
Let $\boldsymbol{x}$ denote the features of a speech clip. For each speech clip, five quality scores are given to quantify five different quality aspects, denoted by $\boldsymbol{y}=(y_1, ..., y_5)$. A multivariate (non-intrusive) speech quality dataset is given by $\mathcal{D}=\{(\boldsymbol{x}_n\, \boldsymbol{y}_n)\}_{n=1}^N$, where $N$ is the total number of speech clips.

Instead of modeling the point estimate $\boldsymbol{y}$, we estimate the multivariate posterior of the MOS for the speech clip $\boldsymbol{x}$. This means we are interested in
\begin{eqnarray}
    p_{\pmb{\psi}}(\boldsymbol{y}|\boldsymbol{x}),
\end{eqnarray}
where $\pmb{\psi}$ are the parameters of the posterior distribution. Following DNSMOS Pro \cite{DNSMOSp}, we model the posterior as a Gaussian motivated by the analytical tractability. Since there are five quality labels, we will model the posterior with a multivariate Gaussian distribution. This means the problem is to estimate $\pmb{\psi}(\boldsymbol{x})=(\boldsymbol{\mu}(\boldsymbol{x}),\Lambda(\boldsymbol{x}))$, since $p_{\pmb{\psi}}(\boldsymbol{y}|\boldsymbol{x}) = \mathcal{N}(\boldsymbol{y};\boldsymbol{\mu}(\boldsymbol{x}),\Lambda(\boldsymbol{x}))$.

Consider a DNN as a regression function $f_{\pmb{\theta}}(\boldsymbol{x})$, with parameters $\pmb{\theta}$, that maps to the Gaussian parameters $\pmb{\psi}(\boldsymbol{x})$; as \begin{eqnarray}
\pmb{\psi}(\boldsymbol{x}) = \boldsymbol{f}_{\pmb{\theta}}(\boldsymbol{x}). 
\end{eqnarray} We can train $\boldsymbol{f}_{\pmb{\theta}}$ in a maximum-likelihood manner using the dataset $\mathcal{D}$. The optimization problem formulation is given by
\begin{eqnarray}
    \arg \max_{\pmb{\theta}} \log \prod_{n=1}^N p_{\pmb{\psi}}(\boldsymbol{y}_n|\boldsymbol{x}_n) = \mathcal{N}(\boldsymbol{y}_n; \pmb{\psi}(\boldsymbol{x}_n)=\boldsymbol{f}_{\pmb{\theta}}(\boldsymbol{x}_n)).
\label{max_likelihood_1}
\end{eqnarray}

\subsection{Architecture}

\begin{figure}[t]
  \centering
  \centerline{\includegraphics[width=8cm]{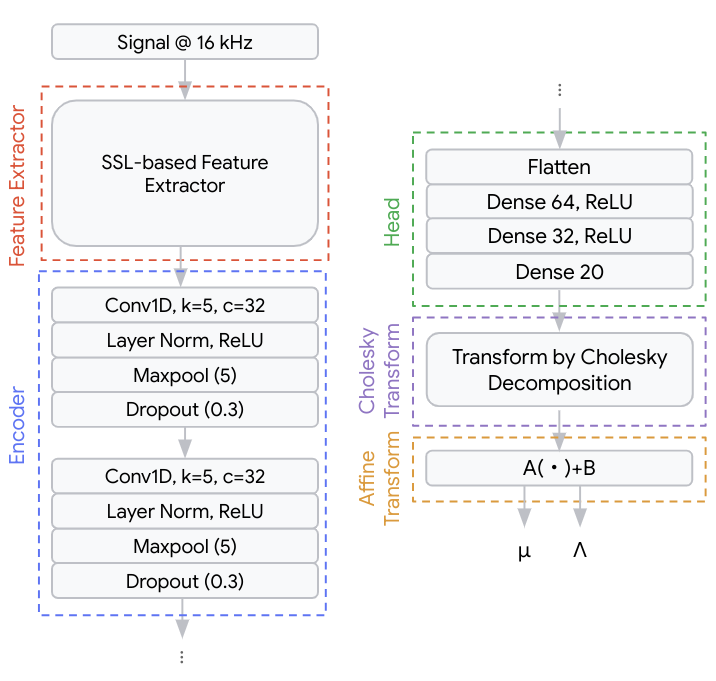}}
\caption{Overview of the architecture.}
\label{fig:arch}
\end{figure}

Following the work of \cite{SSL-MOS}, we use a pre-trained feature extractor, and attach a trainable regressor. Architectural considerations are made to predict a proper covariance matrix. An affine transformation is applied to the output by generalizing the work of \cite{DNSMOSp}. A schematic overview of the architecture is given in Figure \ref{fig:arch}. It consists of five components: \textit{Feature Extractor}, \textit{Encoder}, \textit{Head}, \textit{Cholesky Transform}, and \textit{Affine Transform}.

We adopt wav2vec 2.0 \cite{w2v2} as the feature extractor, selecting layer $12$ based on a preliminary study. A two-layer encoder processes the extracted features using 1D convolutions with a kernel size of $5$, ReLU activations, layer normalization, max pooling of size $5$, and dropout of $0.3$. The head consists of a flattening layer followed by three dense layers with ReLU activations, outputting a 20-dimensional vector for predicting a multivariate Gaussian distribution.

\subsubsection{Cholesky Transform}

A proper covariance matrix in a multivariate Gaussian distribution satisfies two properties: (1) it is symmetric, and (2) it is positive semi-definite; that is, for any $\boldsymbol{z}\in \mathbb{R}^n$, $\boldsymbol{z}^T\Lambda(\boldsymbol{x})\boldsymbol{z}\geq 0$. To the end of predicting a proper multivariate Gaussian distribution, we do a transformation of the output of the head. We call this transformation the \textit{Cholesky transfrom}, since it is based on the Cholesky decomposition explained below. 

The $20$ values from the head are distributed following:
\begin{itemize}
    \item Five values for the mean vector $\boldsymbol{\mu} \in \mathbb{R}^5$.
    \item Fifteen values to construct a lower triangular matrix $L\in \mathbb{R}^{5\times 5}$.
\end{itemize}

By the Cholesky decomposition \cite{cholesky}, a matrix $\Lambda$ is symmetric and positive definite if and only if it can be factorized as $\Lambda=LL^T$, where $L$ is a lower triangular matrix with positive diagonal elements. Subsequently, we define the predicted covariance matrix as  
\begin{equation}\label{eq:cholesky}\Lambda=\textnormal{Softplus}_{diag}(L)\textnormal{Softplus}_{diag}(L^T),
\end{equation}
where $\textnormal{Softplus}_{diag}$ is the Softplus operation, $x\mapsto ln(1+e^x)$, applied only on the diagonal elements.

Cholesky's decomposition ensures that $\Lambda$ is a proper covariance matrix, and that the mapping $L\to \Lambda$ given by Eq. \ref{eq:cholesky} is surjective in the non-trivial case.

\subsubsection{Affine Transform}

The output of the Cholesky transformation, $\boldsymbol{\mu}(\boldsymbol{x})$ and $\Lambda(\boldsymbol{x})$, is further processed by an affine transformation. The transformation is given by \begin{equation}
    (\hat{\boldsymbol{\mu}}(\boldsymbol{x}), \hat{\Lambda}(\boldsymbol{x})) = (A\boldsymbol{\mu}(\boldsymbol{x}) + \boldsymbol{b}, A\Lambda(\boldsymbol{x})A^T),
\end{equation}

where $A=2I$ ($I$ is the unit $5\times 5$ matrix) and $\boldsymbol{b}=(3,3,3,3,3)$.

\begin{table*}

\begin{center}\caption{Performance on NISQA\_VAL\_SIM dataset. For our experiments, the average of $10$ runs is reported, with error margins being one standard deviation of the runs.}
\resizebox{\textwidth}{!}{%
\begin{tabular}{l|cc|cc|cc|cc|cc}
    & \multicolumn{2}{c|}{MOS} & \multicolumn{2}{c|}{NOI} & \multicolumn{2}{c|}{COL} & \multicolumn{2}{c|}{DIS} & \multicolumn{2}{c}{LOUD} \\
    {Model} & {RMSE} & {PCC} & {RMSE} & {PCC} & {RMSE} & {PCC} & {RMSE} & {PCC} & {RMSE} & {PCC} \\ \hline 
    \multicolumn{11}{c}{Not simulated. Results quoted from literature.} \\ \hline
    {NISQA2 \cite{probing}} & $0.532$ & $0.898$ & $0.496$ & $0.871$ & $0.581$ & $0.835$ & $0.619$ & $0.823$ & $0.533$ & $0.807$ \\
    {NISQA59 \cite{NISQA}} & $0.523$ & $0.897$ & $0.497$ & $0.862$ & $0.572$ & $0.809$ & $0.606$ & $0.823$ & $0.523$ & $0.797$ \\
    {Conformer w/ Attn \cite{probing}} & $0.458$ & $0.927$ & $0.477$ & $0.878$ & $0.523$ & $0.860$ & $0.496$ & $0.881$ & $0.527$ & $0.799$ \\ \hline
    \multicolumn{11}{c}{Simulated in our experiments. $10$ runs each.} \\ \hline
    {MultiGauss} & $0.505{\scriptstyle\pm 0.01}$ & $0.922{\scriptstyle\pm 0.00}$ & $0.506{\scriptstyle\pm 0.02}$ & $0.880{\scriptstyle\pm 0.00}$ & $0.532{\scriptstyle\pm 0.02}$ & $0.859{\scriptstyle\pm 0.00}$ & $0.549{\scriptstyle\pm 0.03}$ & $0.876{\scriptstyle\pm 0.00}$ & $0.544{\scriptstyle\pm 0.02}$ & $0.728{\scriptstyle\pm 0.02}$ \\
    {MultiGauss indep} & $0.500{\scriptstyle\pm 0.01}$ & $0.923{\scriptstyle\pm 0.00}$ & $0.503{\scriptstyle\pm 0.01}$ & $0.879{\scriptstyle\pm 0.00}$ & $0.527{\scriptstyle\pm 0.02}$ & $0.858{\scriptstyle\pm 0.00}$ & $0.539{\scriptstyle\pm 0.02}$ & $0.877{\scriptstyle\pm 0.00}$ & $0.548{\scriptstyle\pm 0.03}$ & $0.797{\scriptstyle\pm 0.00}$ \\
    {MultiGauss MSE} & $0.490{\scriptstyle\pm 0.02}$ & $0.922{\scriptstyle\pm 0.00}$ & $0.506{\scriptstyle\pm 0.02}$ & $0.878{\scriptstyle\pm 0.00}$ & $0.521{\scriptstyle\pm 0.01}$ & $0.858{\scriptstyle\pm 0.00}$ & $0.528{\scriptstyle\pm 0.02}$ & $0.877{\scriptstyle\pm 0.00}$ & $0.535{\scriptstyle\pm 0.02}$ & $0.794{\scriptstyle\pm 0.00}$
\end{tabular}}\label{table:performance}
\end{center}
\end{table*}

\begin{table*}
\begin{center}\caption{Performance on NISQA\_VAL\_LIVE dataset. For our experiments, the average of $10$ runs is reported, with error margins being one standard deviation of the runs.}
\resizebox{\textwidth}{!}{%
\begin{tabular}{l|cc|cc|cc|cc|cc}
    & \multicolumn{2}{c|}{MOS} & \multicolumn{2}{c|}{NOI} & \multicolumn{2}{c|}{COL} & \multicolumn{2}{c|}{DIS} & \multicolumn{2}{c}{LOUD} \\
    {Model} & {RMSE} & {PCC} & {RMSE} & {PCC} & {RMSE} & {PCC} & {RMSE} & {PCC} & {RMSE} & {PCC} \\ \hline 
    \multicolumn{11}{c}{Not simulated. Results quoted from literature.} \\ \hline
    {NISQA2 \cite{probing}} & $0.456$ & $0.804$ & $0.524$ & $0.706$ & $0.511$ & $0.532$ & $0.607$ & $0.562$ & $0.528$ & $0.707$ \\
    {NISQA59 \cite{NISQA}} & $0.401$ & $0.822$ & $0.550$ & $0.723$ & $0.454$ & $0.566$ & $0.604$ & $0.542$ & $0.492$ & $0.728$ \\
    {Conformer w/ Attn \cite{probing}} & $0.378$ & $0.861$ & $0.477$ & $0.761$ & $0.484$ & $0.519$ & $0.555$ & $0.618$ & $0.496$ & $0.718$ \\ \hline
    \multicolumn{11}{c}{Simulated in our experiments. $10$ runs each.} \\ \hline
    {MultiGauss} & $0.434{\scriptstyle\pm 0.03}$ & $0.864{\scriptstyle\pm 0.01}$ & $0.510{\scriptstyle\pm 0.02}$ & $0.743{\scriptstyle\pm 0.01}$ & $0.492{\scriptstyle\pm 0.02}$ & $0.567{\scriptstyle\pm 0.02}$ & $0.578{\scriptstyle\pm 0.03}$ & $0.618{\scriptstyle\pm 0.02}$ & $0.532{\scriptstyle\pm 0.02}$ & $0.728{\scriptstyle\pm 0.02}$ \\
    {MultiGauss indep} & $0.430{\scriptstyle\pm 0.02}$ & $0.865{\scriptstyle\pm 0.01}$ & $0.524{\scriptstyle\pm 0.01}$ & $0.728{\scriptstyle\pm 0.02}$ & $0.495{\scriptstyle\pm 0.03}$ & $0.568{\scriptstyle\pm 0.02}$ & $0.567{\scriptstyle\pm 0.02}$ & $0.618{\scriptstyle\pm 0.01}$ & $0.549{\scriptstyle\pm 0.04}$ & $0.719{\scriptstyle\pm 0.02}$ \\
    {MultiGauss MSE} & $0.446{\scriptstyle\pm 0.02}$ & $0.855{\scriptstyle\pm 0.01}$ & $0.528{\scriptstyle\pm 0.02}$ & $0.726{\scriptstyle\pm 0.01}$ & $0.528{\scriptstyle\pm 0.02}$ & $0.541{\scriptstyle\pm 0.02}$ & $0.572{\scriptstyle\pm 0.04}$ & $0.624{\scriptstyle\pm 0.01}$ & $0.539{\scriptstyle\pm 0.04}$ & $0.722{\scriptstyle\pm 0.01}$
\end{tabular}}\label{table:performance_live}
\end{center}
\end{table*}

This transformation transforms the quality labels onto the scale $[-1, 1]$, which removes training bias in the optimization problem formulation \cite{DNSMOSp}. Note that a multivariate Gaussian distribution under an affine transform remains to be a multivariate Gaussian distribution.


\textbf{Output Distribution:} To conclude, the model outputs parameters of a $5$-dimensional multivariate Gaussian distribution:
\begin{equation}
\mathcal{N}(\hat{\boldsymbol{\mu}}(\boldsymbol{x}), \hat\Lambda(\boldsymbol{x})),
\end{equation}
where $\hat{\boldsymbol{\mu}}(\boldsymbol{x})$ is the mean vector and $\hat\Lambda(\boldsymbol{x})$ is the covariance matrix.

\subsection{Training}

Consider a speech quality dataset $\mathcal{D}=\{(\boldsymbol{x}_n\, \boldsymbol{y}_n)\}_{n=1}^N$, where $\boldsymbol{x}_n$ is the features of speech clip $n$ and $\boldsymbol{y}_n$ is the 5-dimensional quality score vector of the speech clip. Training is done on the dataset to maximize the log-likelihood given in \eqref{max_likelihood_1}. Abbreviate $\hat{\boldsymbol{\mu}}_n=\hat{\boldsymbol{\mu}}(\boldsymbol{x}_n)$ and $\hat\Lambda_n=\hat{\Lambda}(\boldsymbol{x}_n)$. Since $\boldsymbol{y}_n$ is modeled by a Gaussian, we have
\begin{equation}
\begin{aligned}
p_{\pmb{\psi}}(\boldsymbol{y}_n|\boldsymbol{x}_n) &= \frac{\exp\left\{-\frac{1}{2}(\boldsymbol{y}_n-\hat{\boldsymbol{\mu}}_n)^T \hat\Lambda_n^{-1} (\boldsymbol{y}_n-\hat{\boldsymbol{\mu}}_n)\right\}}{\sqrt{|\hat\Lambda_n|(2\pi)^{5}}} 
\end{aligned}
\end{equation}
where $\boldsymbol{f}_{\pmb{\theta}}(\boldsymbol{x}_n)=(\hat{\boldsymbol{\mu}}_n, \hat{\Lambda}_n)$.

To maximize the likelihood, we use the Gaussian negative log-likelihood (GNLL) loss to train our model:
\begin{equation}
 \arg \min_{\pmb{\theta}} \sum_{n=1}^N \frac{1}{2}\left[ 
 \log |\hat{\Lambda}_n| + (\boldsymbol{y}_n-\hat{\boldsymbol{\mu}}_n)^T\hat\Lambda_n^{-1} (\boldsymbol{y}_n-\hat{\boldsymbol{\mu}}_n)\right].
\label{loss function}
\end{equation}

\subsection{Inference}

At the time of inference, the model naturally predicts a multivariate Gaussian distribution of the quality scores given a speech clip; for each speech clip feature $\boldsymbol{x}_n$, it predicts $p_{\pmb{\psi}}(\boldsymbol{y}_n|\boldsymbol{x}_n) = \mathcal{N}(\boldsymbol{y}_n;\hat{\boldsymbol{\mu}}(\boldsymbol{x}_n),\hat\Lambda(\boldsymbol{x}_n))$.

For point estimate, we use the maximum likelihood estimator of $\boldsymbol{y}_n$, given by
\begin{eqnarray}
    \hat{\boldsymbol{y}}_n = \max_{\boldsymbol{y}_n} p(\boldsymbol{y}_n|\hat{\boldsymbol{\mu}}(\boldsymbol{x}_n), \hat{\Lambda}(\boldsymbol{x}_n)) = \hat{\boldsymbol{\mu}}(\boldsymbol{x}_n).
\end{eqnarray}

\section{Experiments}
\label{sec:experiments}

\subsection{Datasets}

\begin{table*}
\caption{Average PCC and RMSE across $5$ dimensions for each of $5$ different datasets. For our experiments, the average of $10$ runs is reported, with error margins being one standard deviation of the runs.}
\begin{center}
\resizebox{\textwidth}{!}{%
\begin{tabular}{l|cc|cc|cc|cc|cc}
    & \multicolumn{2}{c|}{NISQA\_VAL\_SIM} & \multicolumn{2}{c|}{NISQA\_VAL\_LIVE} & \multicolumn{2}{c|}{NISQA\_TEST\_LIVETALK} & \multicolumn{2}{c|}{NISQA\_TEST\_FOR} & \multicolumn{2}{c}{NISQA\_TEST\_P501} \\
    {Model} & {RMSE} & {PCC} & {RMSE} & {PCC} & {RMSE} & {PCC} & {RMSE} & {PCC} & {RMSE} & {PCC} \\ \hline 
    \multicolumn{11}{c}{Not simulated. Results quoted from literature.} \\ \hline
    {NISQA2 \cite{probing}} & $0.550$ & $0.865$ & $0.525$ & $0.662$ & $0.829$ & $0.684$ & $0.451$ & $0.865$ & $0.542$ & $0.881$ \\
    {NISQA59 \cite{NISQA}} & $0.545$ & $0.838$ & $0.500$ & $0.676$ & $0.725$ & $0.702$ & $0.528$ & $0.866$ & $0.409$ & $0.877$ \\
    {Conformer w/ Attn \cite{probing}} & $0.496$ & $0.869$ & $0.478$ & $0.695$ & $0.609$ & $0.795$ & $0.385$ & $0.888$ & $0.543$ & $0.884$ \\ \hline
    \multicolumn{11}{c}{Simulated in our experiments. $10$ runs each.} \\ \hline
    {MultiGauss} & $0.527{\scriptstyle\pm 0.02}$ & $0.867{\scriptstyle\pm 0.00}$ & $0.509{\scriptstyle\pm 0.03}$ & $0.704{\scriptstyle\pm 0.01}$ & $0.502{\scriptstyle\pm 0.03}$ & $0.834{\scriptstyle\pm 0.01}$ & $0.458{\scriptstyle\pm 0.02}$ & $0.829{\scriptstyle\pm 0.01}$ & $0.511{\scriptstyle\pm 0.03}$ & $0.862{\scriptstyle\pm 0.01}$ \\
    {MultiGauss indep} & $0.523{\scriptstyle\pm 0.02}$ & $0.867{\scriptstyle\pm 0.00}$ & $0.513{\scriptstyle\pm 0.03}$ & $0.700{\scriptstyle\pm 0.01}$ & $0.488{\scriptstyle\pm 0.03}$ & $0.847{\scriptstyle\pm 0.01}$ & $0.452{\scriptstyle\pm 0.02}$ & $0.827{\scriptstyle\pm 0.01}$ & $0.512{\scriptstyle\pm 0.02}$ & $0.859{\scriptstyle\pm 0.01}$ \\
    {MultiGauss MSE} & $0.516{\scriptstyle\pm 0.02}$ & $0.866{\scriptstyle\pm 0.00}$ & $0.523{\scriptstyle\pm 0.03}$ & $0.694{\scriptstyle\pm 0.01}$ & $0.487{\scriptstyle\pm 0.03}$ & $0.845{\scriptstyle\pm 0.01}$ & $0.457{\scriptstyle\pm 0.03}$ & $0.821{\scriptstyle\pm 0.01}$ & $0.523{\scriptstyle\pm 0.03}$ & $0.854{\scriptstyle\pm 0.01}$
\end{tabular}}\label{table:performance_avg}
\end{center}
\end{table*}

We use the NISQA Corpus \cite{NISQA} for both training and evaluation, which includes two training datasets (NISQA\_TRAIN\_SIM, NISQA\_TRAIN\_LIVE), two validation datasets (NISQA\_VAL\_SIM, NISQA\_VAL\_LIVE), and three test datasets (NISQA\_TEST\_LIVETALK, NISQA\_TEST\_P501, NISQA\_TEST\_FOR). Each speech clip is rated on five quality aspects: overall quality (MOS), noisiness (NOI), coloration (COL), discontinuity (DIS), and loudness (LOUD), with scores ranging from 1 to 5.

NISQA\_TRAIN\_SIM consists of 10,000 English speech clips with five ratings per clip for each of the five rating dimensions, where clean signals are distorted post-recording using additive Gaussian noise, codec artifacts, bandpass filtering, and more. NISQA\_TRAIN\_LIVE contains 1,020 English speech clips with five ratings per clip for each of the five rating dimensions, recorded from live telephone and Skype calls with real distortions. The combined training set comprises 11,020 clips.

For validation, NISQA\_VAL\_SIM includes 2,500 speech clips with five ratings per clip for each of the five rating dimensions, generated similarly to NISQA\_TRAIN\_SIM, while NISQA\_VAL\_LIVE contains 200 speech clips with five ratings per clip for each of the five rating dimensions, recorded under live conditions akin to NISQA\_TRAIN\_LIVE.

The test datasets contain diverse languages and conditions. NISQA\_TEST\_LIVETALK has 232 German speech clips with 24 ratings per clip for each of the five rating dimensions, recorded via smartphones and laptops with real-world distortions. NISQA\_TEST\_FOR comprises 240 Australian English speech clips with approximately 39 ratings per clip for each of the five rating dimensions, featuring simulated and live VoIP distortions (WhatsApp, Zoom, Google Meet, etc.). Similarly, NISQA\_TEST\_P501 consists of 240 British English speech clips with about 28 ratings per clip for each of the five rating dimensions, following the same setup as NISQA\_TEST\_FOR.

\subsection{Experimental designs}

We train three variants of the proposed model, referred to as \textbf{MultiGauss}. The first variant predicts a full five-dimensional Gaussian distribution without output restrictions. The second, \textbf{MultiGauss indep}, assumes statistical independence among the five quality values, predicting only ten parameters (five means and five variances). The third, \textbf{MultiGauss MSE}, predicts only five point estimates using mean-square-error loss instead of a probabilistic distribution. 

\subsection{Results}

We train all models for $30$ epochs using the Adam optimizer with a learning rate of $10^{-4}$, with moving average parameters $\beta_1=0.9$, $\beta_2=0.999$ \cite{Adam}. All signals are downsampled to $16$ kHz and repetitively padded to $8$ s, before processing by wav2vec 2.0. We compare with NISQA2 \cite{probing}, NISQA59 \cite{NISQA}, and Conformer \cite{probing}. The source code is available at: \url{https://github.com/fcumlin/MultiGauss}. 

The performance results on validation data are shown in Table \ref{table:performance} and \ref{table:performance_live}. First, MultiGauss is on par with NISQA2 and Conformer. Secondly, the use of a multivariate Gaussian approach does not inherently degrade performance; on certain dimensions, the probabilistic method outperforms the MultiGauss MSE, leading to improved results.

Furthermore, the MultiGauss model demonstrates consistent performance, with low standard deviation across evaluation metrics and quality dimensions, indicating the robustness of its architecture and optimization objective.

As a generalizability test, we also make inferences on the three out-of-distribution NISQA datasets. The results are shown in Table \ref{table:performance_avg}. As can be seen, the model and variants show high generalizability.

\subsubsection{Importance of affine transformation}

\begin{figure}
\centering
\begin{subfigure}{.25\textwidth}
  \centering
  \includegraphics[width=\linewidth]{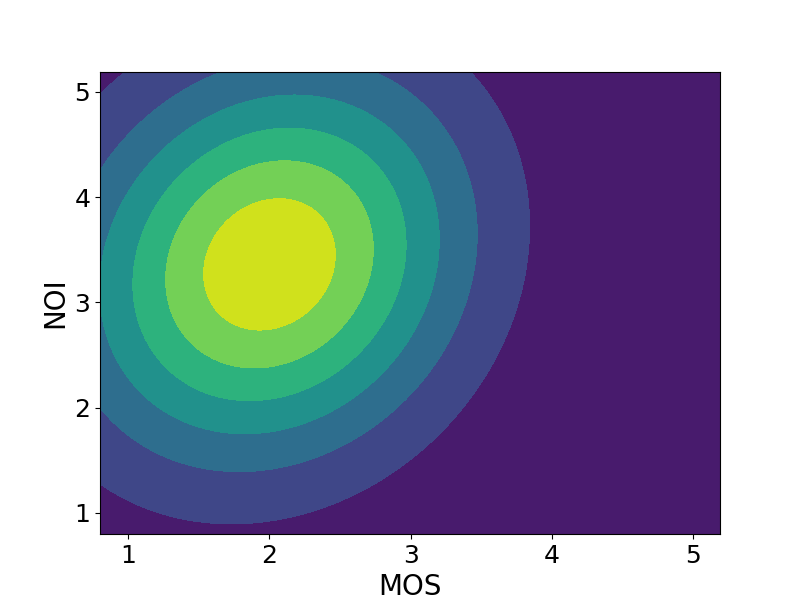}
  \caption{Sample 1}
  \label{fig:marginal_dists_a}
\end{subfigure}%
\begin{subfigure}{.25\textwidth}
  \centering
  \includegraphics[width=\linewidth]{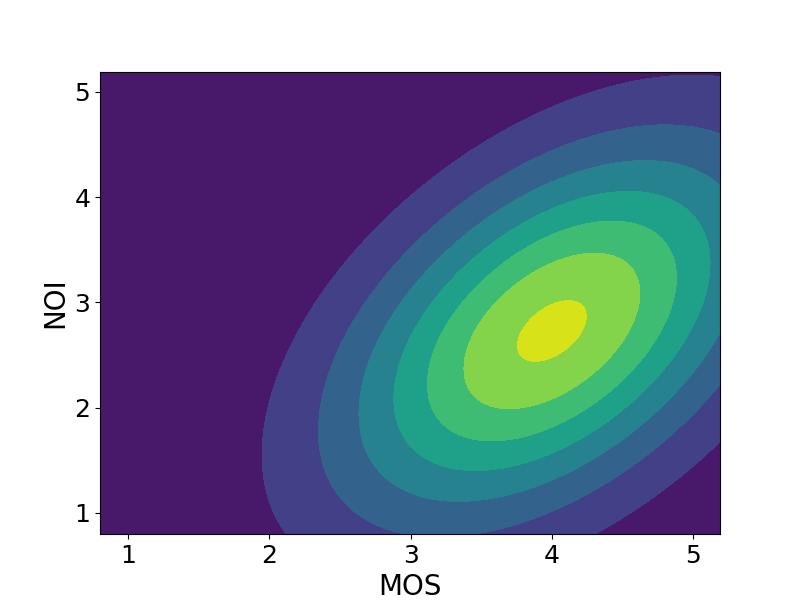}
  \caption{Sample 2}
  \label{fig:marginal_dists_b}
\end{subfigure}
\caption{Contour plots of the marginal distributions over MOS and NOI.}
\label{fig:marginal_dists}
\end{figure}

\begin{figure}
    \centering
    \includegraphics[width=0.8\linewidth]{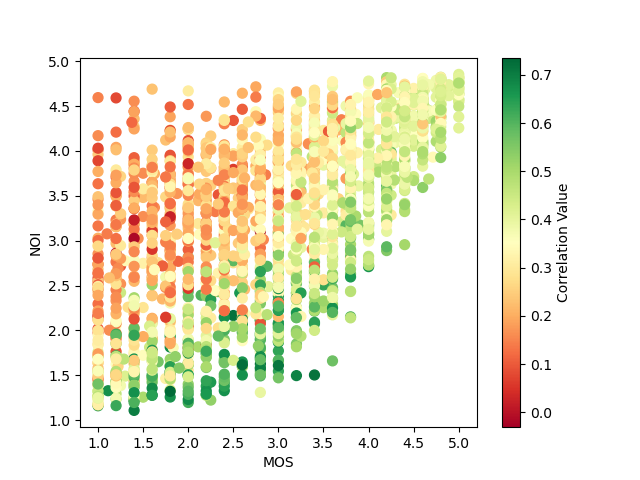}
    \caption{Scatter plot of the MOS and NOI labels on the NISQA\_VAL\_SIM dataset. The correlation values are the predicted correlation values by the MultiGauss model.}
    \label{fig:scatter_with_corr}
\end{figure}

To assess the impact of the affine transformation, we train MultiGauss with and without it. As shown in Table \ref{table:wo_affine}, the transformation significantly reduces RMSE, while having no effect on correlation performance. This aligns with findings in \cite{DNSMOSp}, where the transformation was shown to mitigate training bias.

\begin{table}
\caption{MultiGauss with and without affine transformation.}
\begin{adjustbox}{width=\columnwidth}
\centering
\begin{tabular}{l|cc|cc}
    & \multicolumn{2}{c|}{NISQA\_VAL\_SIM} & \multicolumn{2}{c}{NISQA\_VAL\_LIVE} \\
    {Model} & {RMSE} & {PCC} & {RMSE} & {PCC} \\ \hline 
    {w/ affine} & $0.527{\scriptstyle\pm 0.02}$ & $0.867{\scriptstyle\pm 0.00}$ & $0.509{\scriptstyle\pm 0.03}$ & $0.704{\scriptstyle\pm 0.01}$ \\
    {w/o affine} & $0.864{\scriptstyle\pm 0.06}$ & $0.866{\scriptstyle\pm 0.00}$ & $0.765{\scriptstyle\pm 0.06}$ & $0.706{\scriptstyle\pm 0.01}$ 
\end{tabular}
\label{table:wo_affine}
\end{adjustbox}
\end{table}

\subsubsection{Multivariate prediction}

Due to limited raters, validating multivariate predictions with methods like bootstrapping is difficult. Instead, we will visualize some results.

Fig. \ref{fig:marginal_dists} shows contour plots of predicted marginal distributions for MOS and NOI. First, for poor-quality samples (Fig. \ref{fig:marginal_dists_a}), low covariance between the two dimensions suggests that improving noise may not significantly increase overall quality. Instead, other quality degradations contribute to the poor quality. Second, for low NOI but relatively good overall quality (Fig. \ref{fig:marginal_dists_b}), the overall quality is highly sensitive to noise changes, indicating that reducing noise can significantly improve quality.

Fig. \ref{fig:scatter_with_corr} shows a scatter plot of MOS and NOI labels, with the color gradient representing the predicted correlation between them. The plot reveals a notable pattern: when the speech signal exhibits a poor NOI value, lower than the MOS, the correlation between the two is predicted to be high. This suggests that noise has a significant impact on the perceived quality of the speech signal. Conversely, for signals with a high NOI value but relatively poor overall quality, the correlation is low. This indicates that other factors beyond noise are influencing the overall quality, implying that efforts to reduce noise alone may not substantially improve the perceived quality of these signals.

\section{Conclusion}
\label{sec:conclusion}

We have proposed a novel approach to probabilistic non-intrusive speech quality assessment by predicting a multivariate Gaussian distribution of speech quality scores. Our method generalizes existing maximum likelihood frameworks and affine transformations to a multivariate context. We have made architectural considerations to predict a proper Gaussian distribution, by employing the Cholesky decomposition. The proposed models show stable training and on-par performance with contemporary works, but with the benefit of providing a multivariate posterior distribution of the speech quality.

\bibliographystyle{IEEEtran}
\input{template.bbl}

\end{document}